\documentclass[conference, onecolumn]{IEEEtran}

\usepackage{color}
\usepackage{graphicx}
\usepackage[dvips]{psfrag}
\usepackage{psfrag}
\usepackage{subfigure}
\usepackage{stfloats}
\usepackage{amsfonts}
\usepackage{amsbsy}
\usepackage{amsmath}
\usepackage{amssymb}
\usepackage{comment}
\usepackage{latexsym}
\usepackage{url}
\usepackage{multirow}
\usepackage{rotating}
\usepackage{cite}
\usepackage{dsfont}
\usepackage{color}
\usepackage{nopageno}

\definecolor{utexas}{RGB}{204,85,0}
\definecolor{arsenal}{RGB}{94,26,30}
\definecolor{bar}{RGB}{0,50,166}
\definecolor{tiffany}{RGB}{10,186,181}

\usepackage{widetext}

\begin{document}

\title{ 
Fully Distributed Cooperative Charging  for Plug-in Electric Vehicles in  Constrained Power Networks}

\author{
\IEEEauthorblockN{ M. Hadi~Amini$^{\star}$, Javad~Mohammadi$^{\star}$, Soummya~Kar$^{\star}$\\}
\IEEEauthorblockA{$^{\star}$Department of Electrical and Computer Engineering, Carnegie Mellon University, Pittsburgh, PA\\ \{mamini1, jmohamma, soummyak\}@andrew.cmu.edu\\
}
}

\maketitle
\thispagestyle{empty}

\begin{abstract}
Plug-in Electric Vehicles (PEVs) play a pivotal role in transportation electrification. The flexible nature of PEVs' charging demand can be utilized for reducing  charging cost as well as optimizing the operating cost of power and transportation networks. Utilizing charging flexibilities of geographically spread PEVs requires design and implementation of efficient  optimization algorithms.
To this end, we propose a fully distributed algorithm to solve the PEVs' Cooperative Charging with Power constraints (PEV-CCP). Our solution considers the electric power limits that originate from  physical characteristics of  charging station, such as on-site transformer capacity limit, and allows for containing charging burden of PEVs on the electric distribution network. Our approach is also motivated by the increasing load demand at the distribution level due to additional PEV charging demand. Our proposed approach distributes computation among agents (PEVs) to solve the PEV-CCP problem in a distributed fashion through an iterative interaction between neighboring agents. 
The structure of each agent's update functions ensures an agreement on a price signal while enforcing individual PEV constraints.
 In addition to converging towards the globally-optimum solution, our algorithm ensures the feasibility of each PEV's decision at each iteration. We have tested  performance of the proposed approach using a fleet of PEVs. 


\end{abstract}


\IEEEpeerreviewmaketitle

\section{Introduction}

{\color{black}Transportation electrification induces large electric loading power systems, caused by integration of plug-in electric vehicles (PEVs). This charging load demand, however, can be deployed as a potential source of modifying overall demand profile by shifting flexible PEVs' consumption \cite{Verzijlbergh12}.  Conventional methods for solving the PEV charging coordination problem, referred to as PEV-CC, are mostly  based on centralized control paradigms. These solutions require a decision making entity (PEV aggregator) which calculates and communicates the optimal charging schedules of PEVs based on power system operator's pricing and drivers' needs.  Hence, these methods require large amount of information exchanges between PEVs and aggregators, e.g. in~\cite{Ma2013,Gan13,ParColGraLyg14,Rivera2013,GonAndBoy14,rahbari2014cooperative,alizadeh2014scalable}. This requirement results in increasing the complexity of centralized methods with reduced practicality to optimize a large number of geographically dispersed PEVs~\cite{Bessa2012,Vagropoulos13,Gonzalez14a, hu2014coordinated,sojoudi2011optimal, rotering2011optimal}.  Limitations of legacy infrastructure in accommodating the charging needs of PEVs is a major concern for their integration. For example, ratings of distribution transformers  \cite{geng2013two,amini2017hierarchical} or constraints enforced by demand side management programs, such as peak shaving strategies\cite{shao2011demand},
limit available charging power.}


The structure of the solution depends on the cooperation strategy of agents (PEVs). For instance, non-cooperative agents, considered in \cite{Ma2013,ParColGraLyg14}, utilize mean field game theory for coordinating PEV charging. Along this line, authors in  \cite{Gan13,Rivera2013,GonAndBoy14,rahbari2014cooperative} use internal cooperation among PEVs to solve the charging coordination problem.  Also, alternating direction method of multipliers is used in \cite{Rivera2013,GonAndBoy14} to decompose the original PEV-CC problem into subproblems with less computational complexity.  
However, all above-mentioned methods  need information exchange between an aggregator and PEVs. 

In contrast to centralized algorithms, distributed approaches do not need a central computation unit to optimize the charging schedule of  PEVs. Specifically, consensus-based algorithms (e.g.,  \cite{olfati2007consensus}) have lend themselves as  promising alternative techniques for enabling distributed coordination. They have been widely deployed for various applications, such as load management \cite{asr2013consensus,kar2014distributed},  state estimation \cite{battistelli2015distributed},   and optimal power flow  \cite{Mohammadi_distributedOPF_2014(1)}. In the consensus-based algorithms, agents perform local computations and exchange information with neighboring agents  \cite{dimakis2010gossip,olfati2007consensus} to converge to common optimal solutions, i.e., they hold a copy of coupling variables and reach an agreement on the value of these variables by following an iterative procedure.

A distributed consensus-based approach for the cooperative charging problem of PEVs is proposed in  \cite{rahbari2014cooperative}. In \cite{Mohammadi2015EV,mohammadi2016globsip}, we introduced distributed iterative algorithms of the \emph{consensus}+\emph{innovations} type~\cite{kar2014distributed} to solve the PEV-CC problem. The \textit{consensus} and \textit{innovation} update terms enforce an agreement on a price signal to minimize the charging cost of whole fleet while satisfying the local constraints of the individual PEVs respectively. Further, the \emph{consensus}+\emph{innovations} based Distributed PEV Coordinated
Charging, the  $\mathcal{CI-DPEVCC}$, scheme provided in \cite{mohammadi2016globsip} guarantees feasibility of each PEV's solution at each iteration. The method proposed in this paper, referred to as  the $\mathcal{CI-DPEVCCP}$, i.e., \textit{consensus+innovations} based Distributed PEV Coordinated Charging with Power constraints, extends our previous work presented in \cite{mohammadi2016globsip} by taking into account the power constraints enforced by distribution network limitations while solving the PEV-CC problem. {\color{black} The $\mathcal{CI-DPEVCCP}$ takes into account the global constraint of aggregate charging power limit in a distributed  fashion by proposing a modified update rule as compared with the  $\mathcal{CI-DPEVCC}$. Although adding this new constraint improves the practicality of our model, it changes the nature of the solution  with respect to   \cite{Mohammadi2015EV,mohammadi2016globsip},  that ignored the power constraint. More details on the update rules and the effect of power constraint on  valley-filling capability of PEVs' charging demand are  provided in the following sections.  }

\section{Problem Formulation}

{\color{black}The PEV-CCP problem is concerned with finding the most cost-effective charging schedules for a group of PEVs while satisfying their mobility needs and accommodating  grid constraints, e.g., the capacity limitation of distribution grid transformers\cite{GonAndBoy14}. The  PEV-CCP formulation is provided by}
\small
\begingroup
\allowdisplaybreaks
\begin{align}
\mathrm{minimize}_{\mathbf{x}_{v},\mathbf{L}}&~~c_1  \mathbf{L}^\top \cdot  \mathbf{L}+c_2^\top \cdot \mathbf{L}\label{GeneralPHEVcharging}\\
\textrm{s.t.} ~~\mathbf{L}&=\sum_{v\in V}\mathbf{x}_{v}~~~~~~ \label{loadEq_const}\\
& \mathbf{L}\leq \mathbf{P}_{\text{max}}~~~~~~  \label{power_const}\\
& A\cdot  \mathbf{x}_{v}\leq b_{v}~~\forall v\in\{1,\cdots ,V\} \label{en_const}\\ 
\underline{x}_{v}&\leq \mathbf{x}_{v} \leq \overline{x}_{v}~\forall v\in\{1,\cdots ,V\} \label{pow_const}
\end{align}
\endgroup
\normalsize
{\color{black}where $\mathbf{x}_v$ denotes charging power schedule of PEV $v$ over a given time horizon $[0,T]$, $\mathbf{x}_{v}\in \mathbb{R}^{T\times 1}$, and $\mathbf{L}$ represents aggregated load of PEVs  over a given time horizon $[0,T]$, $\mathbf{L}\in \mathbb{R}^{T\times 1} $. Matrix $A$ and vector $b_{v}$ define the energy constraints of PEV $v$.   Coefficient $c_1\in \mathbb{R}$ and vector $c_2\in \mathbb{R}^{1\times T}$ denote the electricity tariff rates as functions of forecast inflexible load, which does not account for PEV charging demand.  
 Similar to \cite{GonAndBoy14}, the goal of objective function in \eqref{GeneralPHEVcharging} is to minimize the serving costs of both flexible (PEVs) and inflexible loads. Further, we assume that these cost functions are quadratic\footnote{\color{black}Let $\mathbf{L}_\mathrm{in}$ denote  inelastic demand. Hence, cost of serving both elastic and inelastic loads follows the form of $\widetilde{a}\mathbf{1}^\top(\mathbf{L}+\mathbf{L}_\mathrm{in})+\widetilde{b}(\mathbf{L}+\mathbf{L}_\mathrm{in})^\top(\mathbf{L}+\mathbf{L}_\mathrm{in})$, where $\widetilde{a}$ and $\widetilde{b}$ are scalars. Minimizing the objective function in \eqref{GeneralPHEVcharging} with proper values for $c_1$ and $c_2$ is equivalent to minimizing this aggregate cost function of $\mathbf{L}+\mathbf{L}_\mathrm{in}$\cite{Mohammadi2015EV}.}. 
 Moreover,  $\underline{x}_{v}$ and $\overline{x}_{v}$ denote the upper and lower bounds defining the power constraints of an individual PEV $v$. Total number of PEVs is denoted by $V$}. 

{{\color{black}In the proposed formulation, $\mathbf{L}$ is the only coupling variable which appears in \eqref{loadEq_const} and \eqref{power_const}. This variable connects charging decisions of different PEVs. Later in this section, we  derive a distributed representation of \eqref{loadEq_const} and \eqref{power_const} that would fit in our fully distributed set up.}}

{\color{black}Constraint \eqref{power_const} presents the maximum power limit at each time step enforced by transformer capacity limit. According to this constraint, aggregate charging demand of PEVs at each time step is upper bounded by the maximum power limit of the transformer denoted by $P_{\text{max}}$.
 }
 
{{\color{black} Equation \eqref{en_const} presents an abstract representation for energy needs of each PEV based on mobility patterns. This abstract model is derived based on the following model for energy content of the battery at each time step \cite{gonzalez2012centralized}:}}

\begin{equation}
E_{v}(t)=E_{v}(0)+\eta_v\Delta t \sum_{\tau=1}^t \mathbf{x}_{v}(\tau)-\sum_{\tau=1}^t E_{v}^{\text{cons}}(\tau),\label{EnergyEq}
\end{equation}
\normalsize
{{\color{black}Here $E_{v}(0)$ and $E_{v}^{\text{cons}}(t)$ denote the initial energy content of the PEV battery and energy consumption at  time step $t$, respectively. Further,  $\eta_v$ and $\Delta t$ represent the charging efficiency and duration of each time step, respectively. 
The minimum and maximum limits of the battery's energy content are modeled as,}}

\begin{equation*}
 \underline{\mathrm{SOC}}_{v}\leq \frac{E_{v}(t)}{C_v}\leq 1,
\end{equation*}
\normalsize
{
{\color{black}where $\underline{\mathrm{SOC}}_{v}$ denotes the minimum state of charge. Further discussions on the derivation of $A$ and $b$ in \eqref{en_const} are provided in \cite{gonzalez2012centralized}. {\color{black}Note that the elements of matrix $A$ and matrix 
$b$ can be positive, zero, or negative. This prevents the optimization problem from obtaining the trivial solution of all PEVs being charged at their minimum desired power limit, $\underline{x}_{v}$.}
Constraint \eqref{pow_const} presents the upper and lower limits of charging power. Similarly to \cite{gonzalez2012centralized}, we merely consider uni-directional charging of PEVs, i.e., $\underline{x}_{v} = 0$. Further, the upper bound ($\overline{x}_{v}$) is defined as the maximum charging rate of the PEV battery or charging station while a PEV is available for charging, and defined as zero otherwise. Constraints \eqref{en_const} and \eqref{pow_const} only include variables from a single PEV, hence, they are local constraints. Constraints \eqref{loadEq_const} and \eqref{power_const}, however, are global constraints, since they include variables from all PEVs.
The Lagrangian function of the formulated optimization problem is provided by
}
}

\begin{eqnarray}
\mathfrak{L}  &=&  \left(c_1  \mathbf{L}^\top \cdot  \mathbf{L}\right) \nonumber\\
&&  + \lambda^\top\cdot\left( -\mathbf{L}+\sum_{v\in V}\mathbf{x}_{v}\right)\nonumber\\
&&  + \mu_{L}^\top\cdot\left( \mathbf{L}-\mathbf{P}_{\text{max}}\right)\nonumber\\
&& +\sum_{v\in V}\mu^\top_{v}\cdot\left( A\cdot  \mathbf{x}_{v}- b_{v}\right)\nonumber\\
&& + \sum_{v\in V} \mu^\top_{v,-} \cdot\left(\underline{x}_{v}- \mathbf{x}_{v}\right)+  \mu^\top_{v,+}\cdot \left(\mathbf{x}_{v}-\overline{x}_{v} \right)\nonumber,
\end{eqnarray}
\normalsize

{\color{black}
\noindent where $\lambda$'s and $\mu$'s 
denote Lagrange multipliers associated with equality and inequality constraints,  respectively.  First order optimality conditions based on this Lagrangian function are }

\begingroup
 
\allowdisplaybreaks
\begin{alignat}{6}
&&\frac{\partial \mathfrak{L}}{\partial \mathbf{L}} = \left(2c_1 \cdot \mathbf{L}+c_2\right) - \lambda+\mu_{L}=0,&&\label{KKT1}\\
&&\frac{\partial \mathfrak{L}}{\partial \mathbf{x}_{v}} = \lambda+A^\top\cdot \mu_{v}+(\mu_{v,+}-\mu_{v,-})=0&&,\label{KKT2}\\
&&\frac{\partial \mathfrak{L}}{\partial \lambda} = -\mathbf{L}+\sum_{v\in V}\mathbf{x}_{v}=0,&&\label{KKT3}\\
&&\frac{\partial \mathfrak{L}}{\partial \mu_{v}} = A\cdot  \mathbf{x}_{v}- b_{v}\leq0,&&\label{KKT4}\\
&&\frac{\partial \mathfrak{L}}{\partial \mu_{v,+}} =  \mathbf{x}_{v}- \overline{x}_{v}\leq0,&&\label{KKT5}\\
&&\frac{\partial \mathfrak{L}}{\partial \mu_{v,-}} =  -\mathbf{x}_{v}+ \underline{x}_{v}\leq0,&&\label{KKT6}\\&&\frac{\partial \mathfrak{L}}{\partial \mu_{L}} =  \mathbf{L}-\mathbf{P}_{\text{max}}\leq0,&&\label{KKTtrans}
\end{alignat}
\endgroup
\normalsize

{\color{black}\noindent for all $v\in\{1,\ldots,V\}$, as well as  the complementary slackness conditions corresponding to the inequality constraints.} {\color{black}Note that including the power constraint of the aggregate charging demand, that was neglected in \cite{Mohammadi2015EV} and \cite{mohammadi2016globsip}, increases the dimension of the problem by adding \eqref{KKTtrans}  to the KKT conditions. Further, due to the global nature of constraint \eqref{power_const} it increases the complexity of the distributed algorithm, i.e., in contrast with the local constraints of each PEV, it involves variables from all agents as a complicating constraint. In order to tackle the induced complexity, we propose to project the aggregate charging demand at each iteration to  satisfy the upper bound enforced by aggregate power constraint. This is further explained in the next section where we explain the details of our distributed iterative updates. }

\section{Distributed Approach}

{\color{black}In our proposed distributed framework, each PEV is modeled as an agent. The inter-agent communication graph is assumed to be connected, i.e., there is a communication path between every two agents. We further assume that the electricity tariffs ($c_1$ and $c_2$) are available to all agents.

In the proposed $\mathcal{CI-DPEVCCP}$, each agent $v$ updates the local variables, i.e., variables that are directly associated with the PEV  $v$, i.e., $\mathbf{x}_{v}$, $\mathbf{L}_v$, and $\lambda_v$. The iteration counter is denoted by $k$. The update for Lagrange multipliers $\lambda_v$ is given as,}

\begin{eqnarray}
\lambda_{v}(k+1)=\mathbb{P}\left[\lambda_v(k)-\beta_k \overbrace{\left(\sum_{w\in \Omega_v}(\lambda_{v}(k)-\lambda_{w}(k))\right)}^{\text{neighborhood consensus}}\right.\nonumber
-\left.\alpha_k \underbrace{\left(\frac{\mathbf{L}_{v}(k)}{V}-\mathbf{x}_{v}(k)\right)}_{\text{local innovation}}\right]_{[{c_2},\infty)}.\label{LambdaUpdate}
\end{eqnarray}

\normalsize
{\color{black}\noindent where $\alpha_k$ and $ \beta_k$ are positive tuning parameters. Further, $\mathbb{P}$ is the projection operator that enforces $\lambda_{v} \geq c_2$. Note, $\lambda_{v} \geq c_2$ originates from the fact that PEV's electricity consumption can not be a negative value. Hence, the projection operator ensures that the calculated values of $\lambda$ at every iteration meet this condition. 
In \eqref{LambdaUpdate}, the first term accounts for the coupling between the Lagrange multipliers of neighboring agents and ensures the convergence of $\lambda$'s to a (\textit{consensus}) point.
The second term, namely \textit{innovation}, captures the accuracy of each PEV's estimation of the aggregated load ($\mathbf{L}$). If PEV $v$'s charging demand ($\mathbf{x}_{v}$) is exceeding its intuitively expected share of aggregated demand ($\mathbf{L}_{v}(k)/V$), then the value of the \textit{innovation} term increases in the Lagrange multipliers at the next iteration $\lambda_{v}(k+1)$. Given that \eqref{Lupdate} is used in $\mathbf{L}_{v}$ update, PEV $v$'s estimation of total load ($\mathbf{L}_{v}$) increases as the result of $\lambda_{v}(k+1)$ increase.
}

{\color{black}
In order to update the value of $\lambda_{v}$ we directly use \eqref{Lupdate} as follows:}
 
\begin{eqnarray}
\mathbf{L}_{v}(k+1)&=&\mathbb{P}\left[\mathbf{L}_{v}(k)-\frac{1}{2c_1} \frac{\partial \mathfrak{L}}{\partial \mathbf{L}_v(k)}\right]_{(-\infty,{\mathbf{P}_\text{max}}]} \nonumber\\
&=&\mathbb{P}\left[\frac{\lambda_{v}(k)-c_2}{2c_1}\right]_{(-\infty,{\mathbf{P}_\text{max}}]}. \label{Lupdate}
\end{eqnarray}
\normalsize

Inspired by our previous work \cite{Mohammadi_distributedOPF_2014(1)}, the design of the $\mathbf{L}_{v}$ update integrates the maximum power constraint \eqref{power_const} by projecting $\mathbf{L}_{v}$ onto $(-\infty,{\mathbf{P}_\text{max}}]$. This is equivalent to using the full equation (\ref{KKT1}) and including inequality multipliers $\mu_{L}$ to update $\mathbf{L}_{v}$. As the mentioned multipliers only appear in constraint (\ref{power_const}), there is no need to iteratively update them.  {\color{black}This integration technique enables a distributed implementation of (\ref{power_const}) without increasing the complexity of problem caused by larger number of KKT conditions.} Existing methods (e.g., see complicating constraint (2c) in \cite{amini2017hierarchical}) model a transformer's power limit as a complicating constraint that involves variables from all subproblems. In order to solve the resulting problem in a distributed manner, the commonly used approaches decompose of the aforementioned complicating constraint first,  and then update the corresponding Lagrange multiplier at each iteration, which increases their solutions' complexity and run-time. This update makes intuitive sense because we assume that each PEV is aware of the grid's power limitations and take this knowledge into account while updating local variables.

The PEVs' charging schedules are updated according to the following update:
 
\begin{eqnarray}
\mathbf{x}_{v}(k+1)\hspace{-.2cm}&=&\hspace{-.3cm}\mathbb{P}[\mathbf{x}_{v}(k)+\delta_k \left(\frac{\mathbf{L}_{v}(k)}{V}-\mathbf{x}_{v}(k)\right)\nonumber\\
&& -\eta_k\left(\lambda_{v}(k) \right)]_\mathcal{F},\label{Xupdate}
\end{eqnarray}
\normalsize

{\color{black}\noindent where $\delta_k$ and $\eta_k$ are positive tuning parameters. Further, $\mathcal{F}$ defines the (feasibility) region spanned by equations \eqref{en_const} and \eqref{pow_const}. Consequently, the projection operators ensure feasibility of updated values with respect to individual PEV's constraints. In \eqref{Xupdate}, the first term enforces $\mathbf{x}_{v}$ to move towards fulfilling its estimated share of global commitment. The second term of this update rule represents the sensitivity of the Lagrangian function $\mathfrak{L}$ with respect to $\mathbf{x}_{v}$, i.e., $\partial \mathfrak{L}/\partial \mathbf{x}_{v}$. As multipliers $\mu_{v}$, $\mu_{v,+}$ and $\mu_{v,-}$ from $\partial \mathfrak{L}/\partial \mathbf{x}_{v}$ do not appear in any other constraint and the feasibility of the calculated update is ensured by the projection operator, these multipliers are not included in the second term.

Note that the proposed $\mathcal{CI-DPEVCCP}$ algorithm allows for fully distributed implementation of each PEV's update functions since the discussed update rules merely include each agent's corresponding variables and limited information from neighboring agents.}

\section{Simulation Results}
In order to evaluate the performance of the proposed distributed algorithm for cooperative charging of PEVs considering power constraints, we have  conducted the simulations on a fleet of 20 PEVs with maximum charging power of $3.5$kW, efficiency of $0.9$, minimum state of charge of $0.2$, and $C_v$ of either $16$kWh or $24$kWh. Further, the maximum charging power of the whole fleet is $25$kW.

The driving pattern information is derived based on a transportation simulation for Switzerland with the  MATSim software \cite{Balmer2006}, and then represented by $b_v$ (see \cite{gonzalez2012centralized} for more details). A typical winter load in the city of Zurich is considered as the daily load  profile. The optimization horizon is one day, divided into 96 time steps, i.e.,  15-minute intervals. More details about the simulation setup are provided in \cite{mohammadi2016globsip}. We also assume that the communication graph has a ring topology, i.e., each PEV is communicating with exactly two other PEVs. The value of tuning parameters are provided in Table~\ref{TuningValues}. The time-varying tuning parameters are updated at each iteration $k$ as: $\textrm{Tuning Parameter}={\mathcal{r}}/{k^{\mathcal{O}}}$.
According to a  proof provided by \cite{anit2016distributed}, using the above format to update the tuning parameters guarantees the convergence of \textit{consensus+innovations} algorithms. Note  the update in \eqref{LambdaUpdate} is an instance of such algorithms.
 We consider cold start, i.e., initial values of all variables are zero at the first iteration.

\begin{table}[t]

\center
\caption{Tuning Parameter Values}\label{TuningValues}
 
\begin{tabular}{|c|c|c|}
\hline
Parameter & $\mathcal{r}$ & $\mathcal{O}$\\
\hline
$\alpha$ & 10.0222 &0.1600\\
$\beta$ & 0.1080&0.0001\\
$\gamma$ & 0.0080&0.0320\\
$\delta$ & 0.0192&0.0010\\
\hline
\end{tabular}
\end{table}

In order to evaluate the performance of our proposed $\mathcal{CI-DPEVCCP}$ algorithm, we calculate the relative distance of the objective function of the distributed PEV-CCP  at each iteration ($f$) from the optimal value obtained by solving the problem in a centralized fashion ($f^*$) as  $\mathrm{rel}_\mathrm{obj}=\left | f-f^{*} \right |/{f^{*}}$.

\begin{figure}[t]
 \centering
 \includegraphics[width=0.65\textwidth]{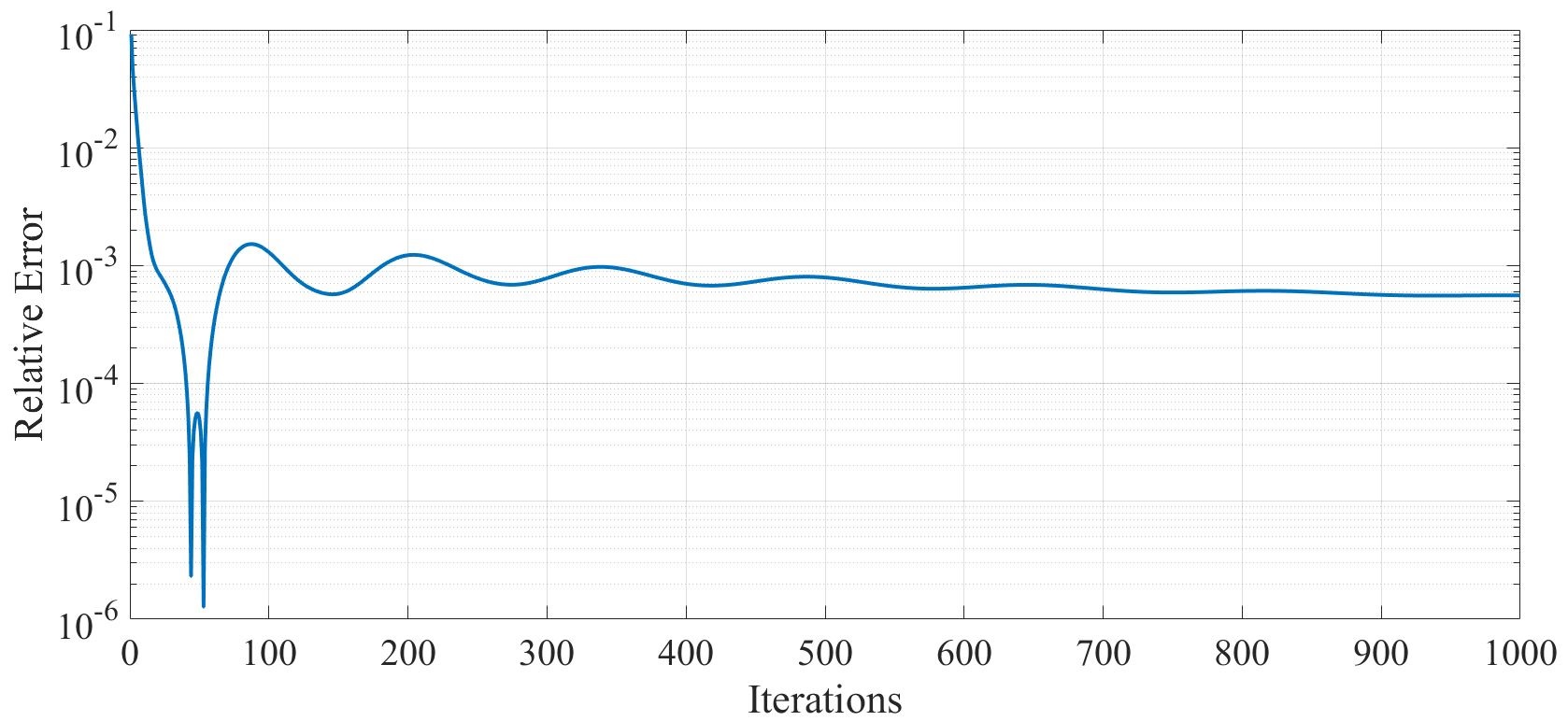}
 \caption{Relative distance from optimal load ($rel_{obj}$), $V=20$.}
 \label{REL-obj-25}
\end{figure}

\begin{figure}[t]
 \centering
 \includegraphics[width=0.65\textwidth]{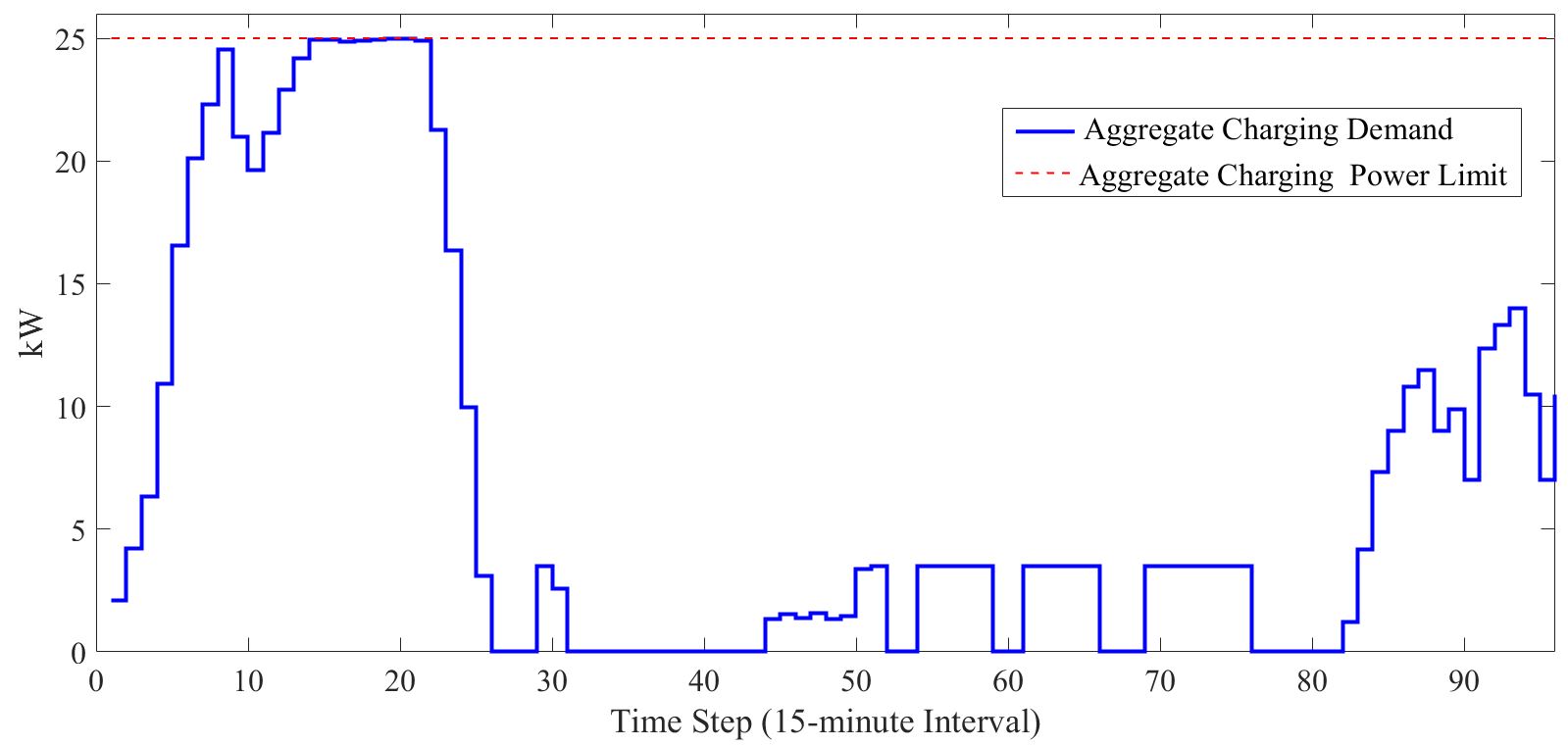}
 \caption{{Aggregate charge demand of PEVs}.}\label{LOADprofile1}
\end{figure}

Figure \ref{REL-obj-25} represents the relative error of the objective function over 1000 iterations. According to this figure, the relative error values converge to $10^{-3}$ after almost 600 iterations.  Oscillations viewed in this figure directly depend on the tuning parameters values. These oscillations could be reduced by adjusting the tuning parameters. Note that this may require a larger number of iterations for convergence. 
Figure~\ref{LOADprofile1}  illustrates the aggregate charge demand determined by our proposed approach. It also verifies that the obtained solution meets the power limit constraint, e.g., $P_{\text{max}}=25$kW.

\begin{figure}[t]
 \centering
  \includegraphics[width=0.65\textwidth]{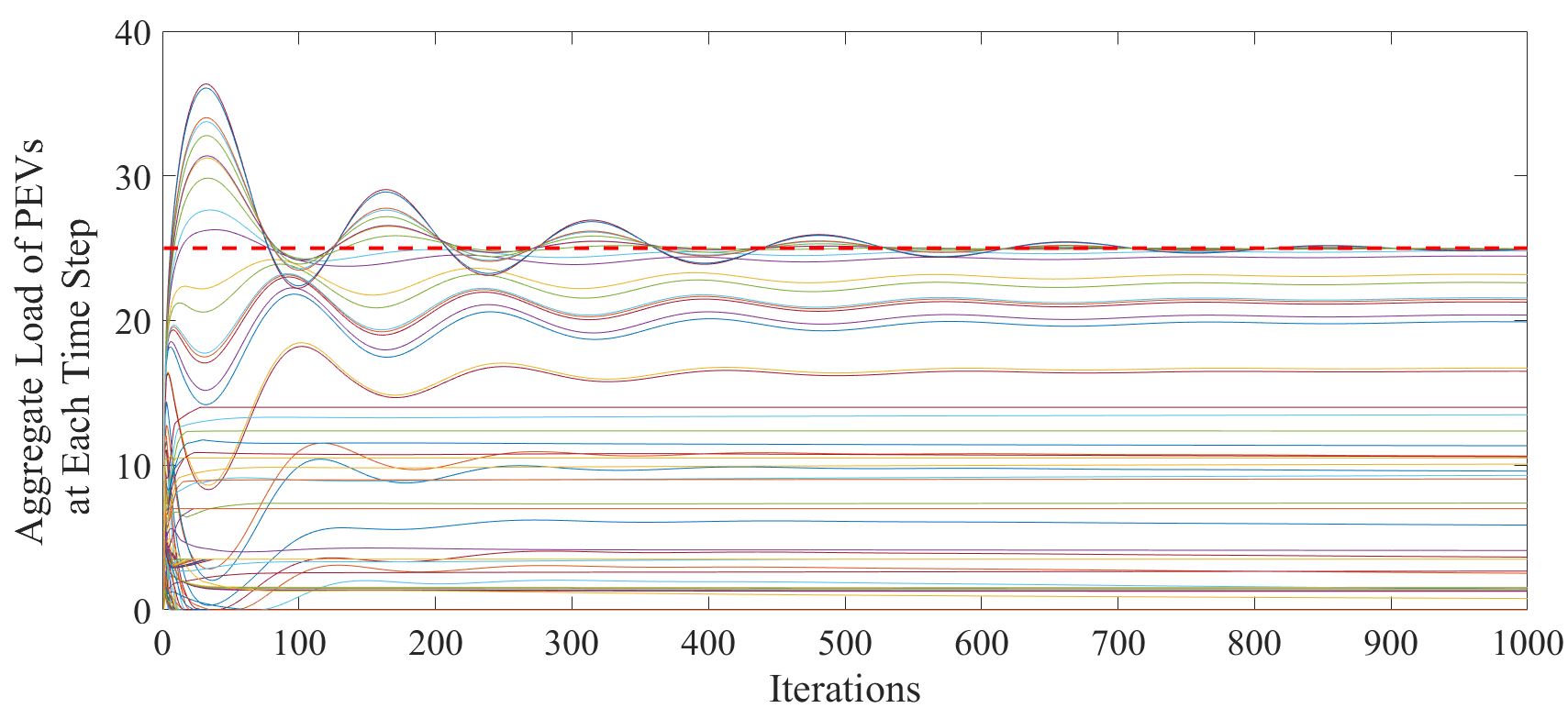}
 \caption{Aggregate PEV loads at each 15-minute time step ($\sum_v \mathbf{x}_{v}^t$), $V=20$, $P_{\text{max}}=25$kW.}
 \label{HourlyLOAD_20}
\end{figure}

\begin{figure}
 \centering
 \includegraphics[width=0.65\textwidth]{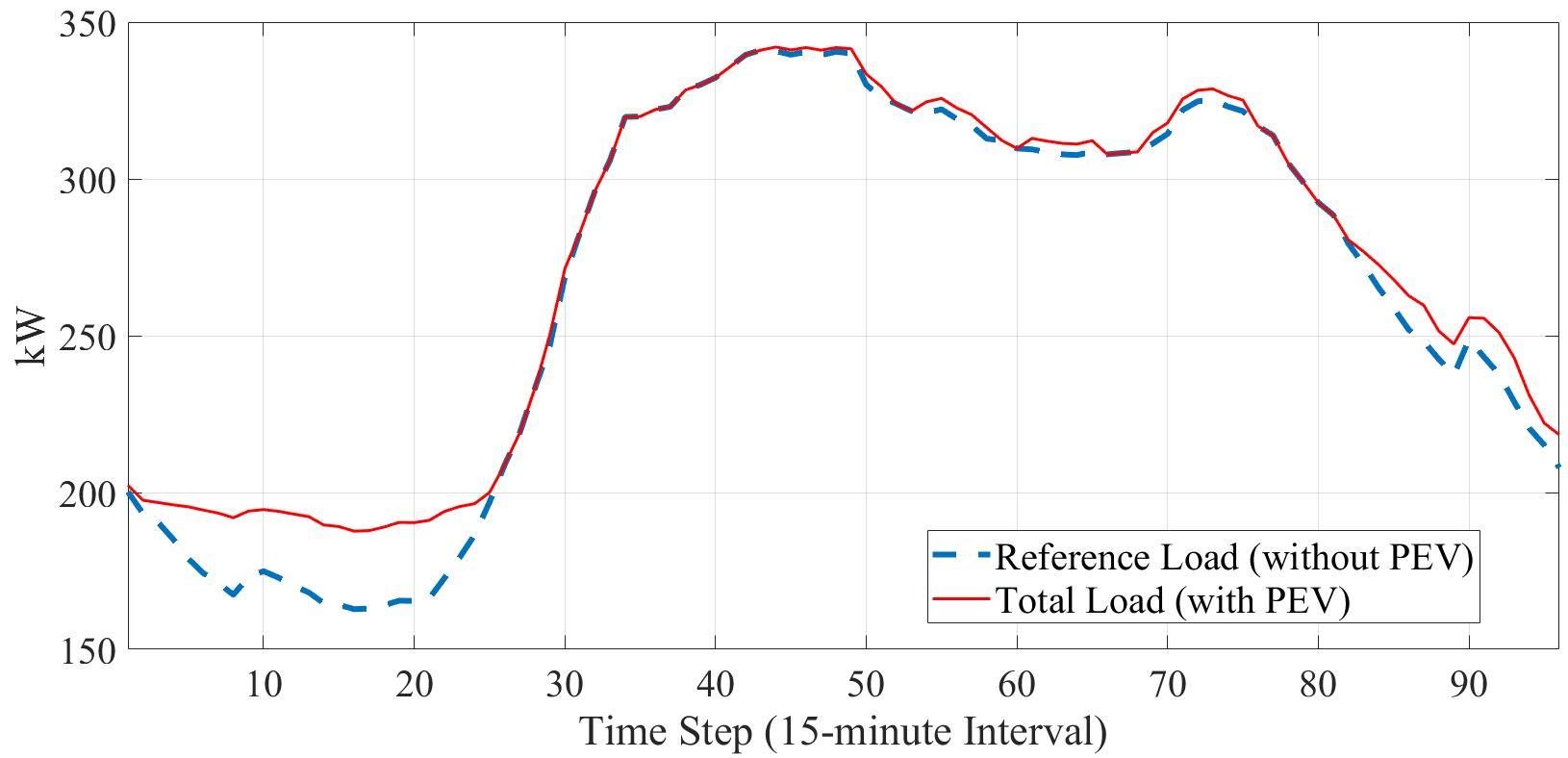}
 \caption{{\color{black}Total load profiles for the full 24 hours horizon}, $V=20$.}\label{LOADprofile}
\end{figure}

Figure ~\ref{HourlyLOAD_20} illustrates the convergence of aggregate load of PEVs for each time step over 1000 iterations. This figure verifies that the $\mathcal{CI-DPEVCCP}$ method guarantees the global constraint (power constraint of the PEV fleet), i.e., the aggregate demand value at each time step is bounded by $P_{\text{max}}$.  Figure~\ref{LOADprofile} illustrates the  load  demand without and with PEV charging when using $\mathcal{CI-DPEVCCP}$. It demonstrates valley-filling capabilities of our proposed solution.  However, unlike our previous approach in \cite{mohammadi2016globsip}, the power constraint limits the contribution of PEVs towards valley-filling which allows for a more realistic implementation.

It is worth noting that each iteration is the $\mathcal{CI-DPEVCCP}$ is not computationally expensive. This is due to the fact that each PEV calculates the algebraic functions provided in \eqref{LambdaUpdate}-\eqref{Xupdate} at each iteration, that can be performed in parallel. Further, note that the obtained solution of each PEV at all iterations leads to a locally feasible solution. 
These feasible solutions, however, might not satisfy the global constraint during the initial iterations, i.e., the power limit constraint is enforced over the iterations. 

\section{Conclusion}
We  proposed a fully distributed \textit{consensus+innovations} based method for the cooperative charging schedule of PEVs considering aggregate charging power constraint.  Our approach optimizes the PEVs' charging cost while satisfying the local constraints of PEVs. Each PEV only needs to communicate with the neighboring agents iteratively. The  update rules lead to locally-feasible solutions of the problem, i.e.,  in case of communication failure the obtained solution at the current iteration can be used as it satisfies local constraints. 
Each agent (PEV) updates its corresponding variables by finding the value of local functions and sharing limited information with the neighbors.  
Unlike the existing methods that  decompose the  complicating constraint corresponding to  power  limit  and then update the corresponding Lagrange multiplier at each iteration, we project the obtained local solution at each iteration to model this constraint. This  reduces the  solution's complexity. 
We have analyzed the proposed algorithm  on a fleet of PEVs to illustrate its  convergence to the optimal solution obtained by centralized solution. {\color{black}Our analysis verifies that the proposed distributed iterative method effectively satisfies the aggregate power constraint. Although prior work demonstrated the significant contribution of PEVs in the valley-filling, we observed that this capability is reduced while considering the power limit. This  observation is expected due to the aggregate power limit constraint that reduces  the flexibility of PEVs, thus changing  the nature of the solution.} 

\section*{Acknowledgment}
 This work was supported in part by the National Science Foundation under Grant Number CCF-1513936, and the Department of  Energy  under Grant Number DE-EE0007165 (SHINES). We would also like to acknowledge the contributions of Prof. Gabriela Hug to this paper. 

\bibliographystyle{IEEEtran}
\bibliography{References1,ReferencesAreaOPF,ReferencesDSEV,literatur}
\vfill

\end{document}